\documentclass[aps,prb,reprint,showpacs,showkeys,superscriptaddress]{revtex4-1}
\usepackage{amssymb,amsmath,amsbsy,graphicx,hyperref,xcolor}

\pdfoutput=1

\begin{document}
\title{Orbital magnetic moments in insulating Dirac systems: Impact on magneto-transport in graphene van der Waals
heterostructures}

\author{Marko M. Gruji\'c}\email{marko.grujic@etf.bg.ac.rs}
\affiliation{School of Electrical Engineering, University of Belgrade, P.O. Box
3554, 11120 Belgrade, Serbia} \affiliation{Department of Physics, University of
Antwerp, Groenenborgerlaan 171, B-2020 Antwerp, Belgium}
\author{Milan \v{Z}. Tadi\'c}\email{milan.tadic@etf.bg.ac.rs}
\affiliation{School of Electrical Engineering, University of Belgrade, P.O. Box
3554, 11120 Belgrade, Serbia}
\author{Fran\c{c}ois M. Peeters}\email{francois.peeters@uantwerpen.be}
\affiliation{Department of Physics, University of Antwerp, Groenenborgerlaan
171, B-2020 Antwerp, Belgium}

\begin{abstract}
In honeycomb Dirac systems with broken inversion symmetry, orbital magnetic
moments coupled to the valley degree of freedom arise due to the topology of
the band structure, leading to valley-selective optical dichroism. On the other hand, in Dirac systems with prominent spin-orbit coupling, similar orbital magnetic moments emerge as well. These moments are coupled to spin,
but otherwise have the same functional form as the moments stemming from
spatial inversion breaking. After reviewing the basic properties of these moments, which are relevant for a whole set of newly discovered materials, such as silicene and germanene, we study the particular impact that these moments have on graphene nano-engineered barriers with artificially enhanced spin-orbit coupling. We examine transmission properties of such barriers in the presence of a magnetic field. The orbital moments are found to manifest in transport characteristics through spin-dependent transmission and conductance, making them directly accessible in experiments. Moreover, the Zeeman-type effects appear without explicitly incorporating the Zeeman term in the models, i.e., by using minimal coupling and Peierls substitution in continuum and the tight-binding methods, respectively. We find that a quasi-classical view is able to explain all the observed phenomena.
\end{abstract}

\maketitle
\section{Introduction}

One of the more intriguing recent developments in the field of graphene
research is the artificial generation of properties that are otherwise
vanishing in intrinsic samples. For instance, carrier mass can be created by
sandwiching graphene with hexagonal boron nitride (hBN), in which case a gap
arises for sufficiently aligned layers.\cite{hunt13,woods14} The occurrence of
the gap is dictated by the interplay of the elastic energy of the graphene
lattice, and the potential energy landscape stemming from hBN.\cite{jung14} The energetically preferred commensurate structure, in which a carbon atom sits on
top of a boron atom, will maximize its area at the expense of other stacking configurations by stretching the graphene layer. This in turn leads to the
appearance of an average gap in the resulting van der Waals heterostructure.\cite{woods14}

On the other hand, it was postulated that spin-orbit coupling (SOC) in graphene can be enhanced by hydrogen adsorption, which forces local rehybridization of
bonds.\cite{neto2009} Note that quantum spin Hall transport signatures introduced by random adatoms are well described by models taking into account a renormalized and uniform SOC.\cite{shevtsov12,jiang12} Moreover, the proximity
effect caused by an appropriate substrate was speculated to lead to SOC
enhancement as well. Both of these mechanisms were recently confirmed
experimentally, opening new avenues for theoretical research.\cite{ozy13,avsar14}

While in graphene the carrier mass and SOC have to be artificially engineered, they are ubiquitous in other group IV monolayers such as silicene, germanene, and
stanene, thanks to their buckled structure and the heavier constituent
atoms.\cite{cahangirov09,liu11,xu13} Given their honeycomb
lattice, they also belong to the same class of materials as graphene, with relativistic
quasiparticles described by the Dirac equation. From the
theoretical point of view, both of the aforementioned parameters appear in a similar
form in the low-energy continuum picture. They are captured by staggered
potential terms $\Delta$ and $\Delta_{SO}$ in the case of mass and SOC,
respectively.\cite{footnote} The term "staggered potential" originates in the language of the
tight-binding method, and it refers to the breaking of the sublattice symmetry by
a traceless potential. Unlike SOC, for which the staggered potential changes
sign depending on the spin and valley of the electron, $\Delta$ opens up a
topologically trivial band gap in the vicinity of the $K$ and $K^{\prime}$
points through inversion symmetry breaking.\cite{kane05a}

At the same time, however, the inversion symmetry breaking leads to a nontrivial alteration of the semiclassical equations of motion on a honeycomb
lattice.\cite{chang95,xiao07} The quantum corrections, which reflect the impact of the Berry phase, and are therefore topological in nature, are twofold. On
the one hand, when subjected to an electric field in the plane, massive Dirac
fermions will attain a velocity component transverse to the field, which is
opposite in the two valleys, thus giving rise to the valley Hall effect. This
effect was recently observed experimentally in a ${\rm MoS}_2$ device, as well
as in graphene-hBN heterostructures.\cite{mak14,gorbachev14} On the other hand, self-rotation of electron wavepackets near the two valleys will produce
valley-contrasting orbital magnetism.\cite{xiao07}

It is well established that the valley Hall and intrinsic spin Hall effects
share the same origin, reflecting the Berry curvature properties of the
underlying insulating systems, generated by $\Delta$ and $\Delta_{SO}$ terms, respectively. Therefore, the two Hall effects are fully
analogous.\cite{feng12,xiao12} Valley-contrasting magnetism was first reported
in Ref.~\onlinecite{xiao07}. On the other hand, we recently found evidence of the corresponding spin-contrasting magnetism in transport calculations involving spin-orbit barriers in bulk graphene,\cite{grujic14} which motivated us to explore the subject more thoroughly. These orbital magnetic moments were previously investigated in a more generalized analysis on the abundance of Hall effects (and the accompanying wealth of orbital magnetization), found in multilayer graphene systems in Ref.~\onlinecite{zhang11}. There, the electron-electron interaction leads to various broken symmetry phases, denoted by the general term "pseudospin ferromagnetism",\cite{min08,zhang10,nand10,jung11} which are captured with a diverse set of mass terms in the low-energy continuum approximation, in models analogous to the ones studied in this paper.

This emerging orbital magnetism is a mechanism that effectively alters the Zeeman energy, and it is the subject of this paper, particularly the moments associated with spin-orbit coupling in monolayer Dirac systems. We first review how the intrinsic SOC in honeycomb monolayers gives rise to orbital magnetic moments coupled to spin, in the same way in which inversion symmetry breaking gives rise to moments coupled to the valley degree of freedom. These moments are completely analogous in nature, and they share exactly the same functional form, apart from coupling to different degrees of freedom. We derive expressions for the moments using both tight-binding and continuum theories, and we show their impact on the Landau level (LL) quantization in the presence of a magnetic field.

Finally, we investigate the influence of the moments on the magneto-transport properties, where we look at the transmission through a barrier with enhanced spin-orbit coupling in graphene. Such a barrier could be realized by an appropriately formed van der Waals heterostructure in an otherwise fully ultrarelativistic material.\cite{shevtsov12,jiang12,ozy13,avsar14} We discuss this case in great
detail from the semiclassical point of view, and we present conclusions that are of practical relevance, namely how the device conductance is affected by orbital
magnetism. In the end, we show that the results are identical whether one uses
the continuum Dirac theory or the tight-binding nonequilibrium Green function
method (TB NEGF) when calculating the transport properties. Remarkably, both approaches yield Zeeman-type transport signatures while employing the magnetic field only through kinetic terms, without actually enforcing the coupling of the spin with the magnetic field, which reflects the orbital nature of the magnetic moments.

\section{Orbital moments in the tight binding picture}\label{II}

\begin{figure}
\centering
\includegraphics[width=8.6cm]{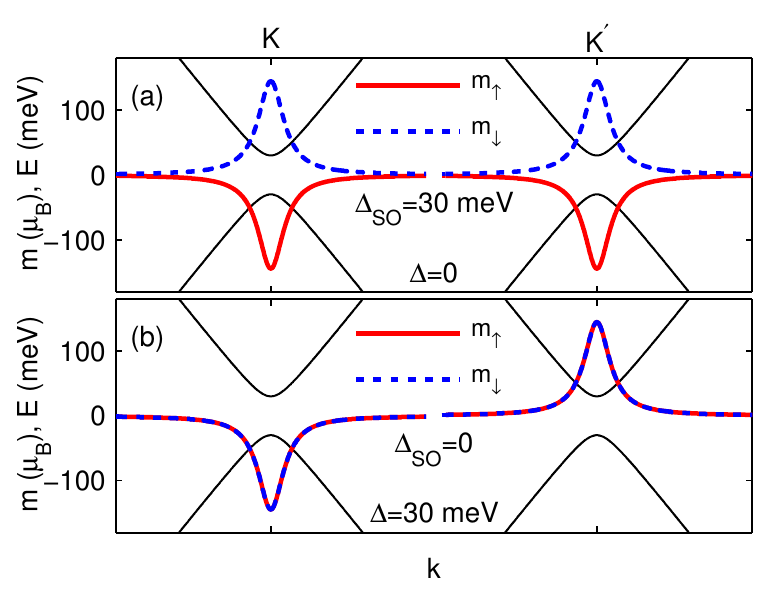}
\caption{The orbital magnetic moments of the spin-up (spin-down) states shown by
thick red (dashed blue) lines, and the corresponding low-energy band structure,
shown in black, for: (a) $\Delta=0$ and $\Delta_{SO}=30$ meV, and (b)
$\Delta=30$ meV and $\Delta_{SO}=0$. Note that in (b) the orbital magnetic
moments for the two spins are equal, due to the absence of SOC.}
\label{0}
\end{figure}

We start with the low-energy tight binding Kane-Mele Hamiltonian valid for a whole set of Dirac materials with prominent intrinsic spin-orbit coupling,\cite{kane05}
\begin{equation}\label{dirac-weyl0}
H=\hbar v_F\left[\tau k_x\sigma_x+k_y\sigma_y\right]+s\tau\Delta_{SO}\sigma_z,
\end{equation}
where $v_F$ is the Fermi velocity, $\Delta_{SO}$ is spin-orbit coupling,
$\sigma_z$ is a Pauli matrix operating in the sublattice subspace, $s=+1/-1$
labels the spin $\uparrow/\downarrow$, and $\tau=+1/-1$ labels the valley
$K/K^{\prime}$. As already mentioned this form of SOC is universal to all group IV monolayers other than graphene, in which on the other hand it could be artificially generated. Note that here $k_x$ and $k_y$ are only parameters, and
not operators. The dispersion relations extracted from Eq. \eqref{dirac-weyl0}
are shown by solid black curves in Fig. \ref{0}(a).

The Hamiltonian \eqref{dirac-weyl0} describes a two-state, electron-hole
symmetric system. For such systems, the orbital magnetic moment ($m$) is
directly proportional to the Berry curvature ($\Omega$),
$m\sim\Omega$.\cite{xiao07,zhang11,yao08} On the other hand, the system is also time-reversal invariant,
and since we disregard the staggered potential $\Delta$ at the moment,
inversion symmetry is not broken either. Since for spatial-inversion and
time-reversal symmetric systems Berry curvature vanishes,\cite{xiao10,chang08}
one might conclude that the orbital moments must vanish as well. However, it is rarely stressed that this only holds for {\it spinless} electrons, which is not the case considered here.\cite{gradhand12} In fact, the Hamiltonian
\eqref{dirac-weyl0} describes a topological insulator, having a non-zero and
opposite Chern number for opposite spins.\cite{kane05} This is because the Kane-Mele model is formed by two opposite copies of the Haldane model,\cite{haldane88} thus breaking the time-reversal symmetry
separately in each spin sector. Since the Chern number is obtained as an
integral of $\Omega$ over the Brillouin zone, the Berry curvature is
nontrivial, and consequently, the orbital magnetic moments will be nonzero.

The orbital moments are perpendicular to the monolayer, and originate from the self-rotation of the electron wave
packet around its center of mass, and they can be obtained from the tight binding
Bloch eigenfunctions $|u\left(\mathbf{k}\right)\rangle$\cite{xiao07,xiao10,zhang11}
\begin{equation}
m=-i\frac{e}{2\hbar}\langle\mathbf{\nabla}_{\mathbf{k}}u|\times\left[H-E\left(\mathbf{k}\right)\right]|\mathbf{\nabla}_{\mathbf{k}}u\rangle,
\end{equation}
which makes their topological origin much clearer. For the particular
Hamiltonian in Eq. \eqref{dirac-weyl0}, we have
\begin{equation}
|u\left(\mathbf{k}\right)\rangle=\left(\begin{array}{c}\sqrt{\frac{E+s\tau\Delta_{SO}}{2E}}\\ \tau\sqrt{\frac{E-s\tau\Delta_{SO}}{2E}}e^{i\tau\phi}\end{array}\right),
\end{equation}
where $E$ is the electron energy, and $\phi=\arctan k_y/k_x$. It is then
straightforward to show that the expression for the magnetic moments that
arise from the spin-orbit coupling reads
\begin{equation}\label{orbitalmagmom}
m=-s\frac{e\hbar v_F^2\Delta_{SO}}{2\left(\Delta_{SO}^2+\hbar^2v_F^2k^2\right)}.
\end{equation}
Variations of the orbital moments in the vicinity of the Dirac points are shown for both spins in Fig. \ref{0}(a). They are maximum near the band edges, decay
away from the two Dirac points, and are obviously opposite for opposite spins.

One can compare these moments with the valley-contrasting moments, arising for
$\Delta_{SO}=0$ and $\Delta\neq0$.\cite{xiao07,xiao10} Their magnitude is given by
\begin{equation}\label{orbitalmomdol}
m=-\tau\frac{e\hbar v_F^2\Delta}{2\left(\Delta^2+\hbar^2v_F^2k^2\right)},
\end{equation}
and they are depicted in Fig. \ref{0}(b). It is clear that the two sets of moments share a similar functional form, except the former couple to spin,
while the latter couple to the valley degree of freedom.\cite{zhang11} The energy region where the moments are prominent was termed the Berry curvature hot spot in Ref. \cite{gorbachev14} There it was unequivocally shown that the gap in well-aligned graphene-hBN van der Waals heterostructures is accompanied by the introduction of nontrivial Berry curvature.

Finally, in the case of both nonzero $\Delta_{SO}$ and $\Delta$, and in the low energy limit, the magnetic moment is given by
\begin{equation}\label{orbitalmagmom0}
m=-\frac{e\hbar v_F^2}{2\left(s\Delta_{SO}+\tau\Delta\right)}.
\end{equation}

The orbital magnetic moments are responsible for the optical selection rules of light absorption in Dirac materials, through the so-called circular dichroism
effect.\cite{yao08,xiao12,ezawa12} Note that the orbital moments in Eq.
\eqref{orbitalmagmom} can dominate the Zeeman response of a system, since they
can be orders of magnitude stronger than the free-electron Bohr magneton for
realistic SOC strengths found in typical Dirac materials.\cite{xiao07,xiao10,zhang11,jung11} In other words, they
will lead to a renormalization of the Land\'{e} $g$ factor, which was recently
observed for transition metal dichalcogenides from first-principles calculations.\cite{kormanyos14}

\section{Landau levels, pseudospin polarization and orbital moments in the continuum picture}\label{III}
\subsection{Landau levels}

We proceed with the case of an applied perpendicular magnetic field ${\bf
B}=B{\bf e}_z$ in bulk graphene, which is included in the Hamiltonian through minimal coupling
\begin{equation}\label{dirac-weyl}
H=\hbar v_F\left[\tau k_x\sigma_x+(k_y+\frac{e}{\hbar}A_y)\sigma_y\right]+s\tau\Delta_{SO}\sigma_z+\Delta\sigma_z.
\end{equation}
This equation could be employed to solve the electron spectrum in the Dirac
system in the presence of $\Delta_{SO}$, $\Delta$, and magnetic field. It will
subsequently lead us to resolve the magnetic moments. Here, the Landau gauge
$\mathbf{A}=\left(0,A_y\right)$ with $A_y=Bx$ is chosen. In this gauge, $k_y$ is a good quantum number and the solutions have the form
$\Psi(x,y)=\exp(ik_yy)\left(\psi_A(x),\psi_B(x)\right)^T$. Introducing $\hbar
v_F\epsilon=E$, $\hbar v_F\delta=s\tau\Delta_{SO}+\Delta$, one can obtain the
LLs in the infinite graphene sheet. In solving the LL spectrum it is useful to
adopt the operators $b_{\tau}^{\dagger}=-i(l_B/\sqrt{2})\left(\tau
k_x+ik_y+ieA_y/\hbar\right)$ and $b_{\tau}$, where $l_B=\sqrt{\hbar/eB}$
denotes the magnetic length. $b_\tau^{\dagger}$ and $b_\tau$ are the bosonic
ladder operators, and they satisfy
$\left[b_{\tau},b_{\tau}^{\dagger}\right]=\tau$. It could be useful to define
these operators such that they fully correspond to the standard ladder
operators of the quantum harmonic oscillator (QHO) shifted by $x_0=k_yl_B^2$
and having the mass $m=\hbar^2/l_B^4k$. Then the eigenstates will be given by
the standard (obviously shifted and rescaled) QHO solutions
\begin{equation}\label{qho}
\langle x|n\rangle=\frac{1}{\sqrt{2^nn!}}e^{-\left(x/l_B+k_yl_B\right)^2/2}H_n\left(\frac{x}{l_B}+k_yl_B\right),
\end{equation}
where $H_n$ are Hermite polynomials. The problem can now be solved in terms of
these solutions for the case of the regular two-dimensional (2D) electron gas in a magnetic
field, having in mind that $b_1^{\dagger}|n\rangle=\sqrt{n+1}|n+1\rangle$,
$b_1|n\rangle=\sqrt{n}|n-1\rangle$ and $b_1|0\rangle=0$, and that the ladder
operators change character in the $K^{\prime}$ valley. The system of coupled
equations with ladder operators is now given by
\begin{align}
\delta\psi_A-i\frac{\omega_c}{v_F}b_{\tau}\psi_B&=\epsilon\psi_A,\\
i\frac{\omega_c}{v_F}b_{\tau}^{\dagger}\psi_A-\delta\psi_B&=\epsilon\psi_B,
\end{align}
where $\omega_c=\sqrt{2}v_F/l_B$ is the cyclotron frequency for Dirac-Weyl
electrons. Then for $n\geq1$ the energies of the LLs are given by
\begin{equation}\label{eq:Landaun}
\epsilon_{n,s,\tau,\pm}=\pm\sqrt{\delta^2+n\omega_c^2/v_F^2}.
\end{equation}
The $s$ and $\tau$ quantum numbers are contained implicitly in the definition
of $\delta$. The joint spinor for the two valleys can be written as
\begin{equation}\label{spinorn}
|n,s,\tau,\pm\rangle=\left(\begin{array}{c}|n-\frac{\tau}{2}-\frac{1}{2}\rangle\\ \pm i\left[\frac{\omega_c\sqrt{n}}{\left(\epsilon+\tau\delta\right)v_F}\right]^{\tau}|n+\frac{\tau}{2}-\frac{1}{2}\rangle\end{array}\right).
\end{equation}

The case of $n=0$ needs special attention, since the solution, Eq.
\eqref{spinorn}, is not valid in this case. Then, the appropriate choice for
the solution is
\begin{equation}\label{spinor0}
|0,s,\tau\rangle=\left(-\tau/2+1/2,\tau/2+1/2\right)^T|0\rangle,
\end{equation}
while the energies are expressed as \cite{tabert13}
\begin{align}\label{eq:Landau0}
\epsilon_{0,s,\tau}=-\tau\delta.
\end{align}
It is worth pointing out however, that observation of the conductance plateaus corresponding to the derived spectrum can depend on the symmetry class of the disorder present in the samples.\cite{ostrovsky08} Note also that these eigenvectors and eigenvalues reduce to the ones for the massless fermions, under the requirement $\delta\rightarrow0$, collapsing the level \eqref{eq:Landau0} to zero energy. Nevertheless, massless fermions can also display quantum Hall signatures, such as a magnetic-field-independent plateau at zero filling factor, which originate from valley mixing scattering processes.\cite{ostrovsky08}

\begin{figure}
\centering
\includegraphics[width=8.6cm]{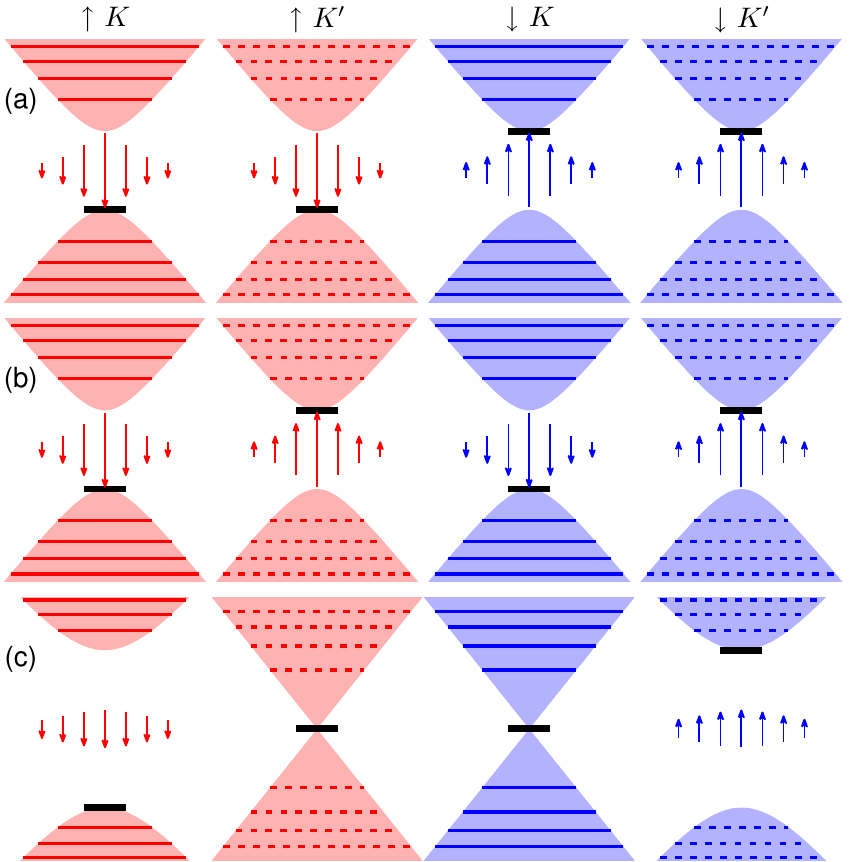}
\caption{Several lowest Landau levels of all spin and valley flavors for (a) $\Delta_{SO}=30$ meV and $\Delta=0$, (b) $\Delta_{SO}=0$ and $\Delta=30$ meV, and (c) $\Delta_{SO}=\Delta=30$ meV. The $n=0$ Landau level is depicted by the horizontal solid black line. Also shown are the bulk bands as red (spin up) and blue (spin down) shaded regions, as well as the sketch of the corresponding orbital moments, with the length of the arrow being proportional to the intensity.}
\label{11}
\end{figure}

Thus the SOC and mass terms split and shift the zeroth LLs away from zero
energy as depicted in Fig. \ref{11}, which shows the low-lying Landau levels for (a) only $\Delta_{SO}\neq0$, (b) only $\Delta\neq0$ and (c) $\Delta_{SO}=\Delta\neq0$ at $B=2$ T. We also depict the bands, as well as the emerging magnetic moments, given by Eqs. \eqref{orbitalmagmom} and \eqref{orbitalmomdol}. The orientation of the moments is related to the position of the $n=0$ Landau level, which is shown by the horizontal solid black lines. Note that the zeroth LLs always reside
on the edges of the appropriate bands. The duality $\Delta_{SO}\leftrightarrow\Delta$, $s\leftrightarrow\tau$ present
in Eq. \eqref{eq:Landau0} is evident in Fig. \ref{11}. In other words, SOC couples the LLs to spin in the
same way that mass couples them to the valley degree of
freedom.\cite{tabert13,krsta12,lado13} The state depicted in Fig. \ref{11}(c) is dubbed spin-valley-polarized metal,\cite{ezawa12a} and it hosts both a massless (lacking the orbital moments) and a massive relativistic Landau spectrum. It can appear in silicene subjected to a perpendicular electric field, for instance. On the other hand, in transition-metal dichalcogenides, both parameters are inherently present, with $\Delta>\Delta_{SO}$, and SOC splits only the LLs in the valence band, yielding a unique set of Hall plateaus.\cite{li13}

\subsection{Orbital moments}

The underlying explanation for the behavior of the LL spectrum can be sought in the existence of orbital magnetic moments.\cite{xiao07,cai13,koshino10} In a
similar fashion to Ref.~\onlinecite{koshino10}, we can obtain the effective
Bohr's magneton in the presence of $\Delta_{SO}$, starting from the Dirac-Weyl
equation, and expanding near the conduction band bottom. We first point out
that near the bottom of the conduction bands, the sublattice pseudospins get
polarized perpendicular to the graphene sheet, with the majority of the weight
concentrated on the A (B) sublattice for $\delta>0$ ($\delta<0$). Likewise, at
the top of the valence band, most of the weight is found on the A (B) sublattice for $\delta<0$ ($\delta>0$). This is obvious for the zeroth LLs, and it occurs in the $\delta=0$ limit as well,\cite{goerbig11,grujic11} but to see it
for higher levels it is helpful to derive the expectation value for the
sublattice pseudospin,
\begin{equation}\label{pseudospinz}
\langle n,s,\tau,\pm|\sigma_z|n,s,\tau,\pm\rangle=\frac{\delta}{\epsilon}.
\end{equation}
which is exactly the same as in the absence of SOC and magnetic
field,\cite{leyla11} only now it is to be used for the discrete energy values
corresponding to the Landau levels. Therefore,
perfect pseudospin polarization is achieved in the bottom (top) of the
conduction (valence) band.

On the other hand, decoupling the Dirac equation gives
\begin{equation}\label{difjna}
\left[k_x^2+\left(k_y+\frac{x}{l_B^2}\right)^2\pm\frac{\tau}{l_B^2}\right]\psi_{A/B}=\left(\epsilon^2-\delta^2\right)\psi_{A/B}.
\end{equation}
Therefore, there is a spatially uniform term proportional to the magnetic
field, with opposite signs on opposite sublattices and opposite valleys.
Consider the importance of this term for states whose sublattice pseudospin
mostly lies in the graphene plane, i.e. for states far away from the band gap,
Eq. \eqref{pseudospinz}. For such states, the two signs play a tug of war,
effectively canceling each other out. However, near the band gap, sublattice
polarization occurs, and the term corresponding to a majority sublattice starts dominating over the other, giving rise to an effective paramagnetic moment. For instance, when $\delta>0$ sublattice A dominates for low electron energies, and the upper sign starts impacting the electron motion. To fully appreciate this
fact, and to write the equation in a manifestly paramagnetic form, one needs to perform a low energy expansion for the equation of the majority sublattice.
After reintroducing $E$, $\Delta$, and $\Delta_{SO}$ explicitly, we can write
$E=\xi+\left(s\tau\Delta_{SO}+\Delta\right)$ for $\delta>0$, and
$E=\xi-\left(s\tau\Delta_{SO}+\Delta\right)$ for $\delta<0$. Taking the limit
$\xi\rightarrow0$, the following equation is obtained for the bottom of the conduction band
\begin{equation}
\left[\frac{p_x^2}{2m_{eff}}+\frac{\left(p_y+eA_y\right)^2}{2m_{eff}}+\frac{e\hbar v_F^2B}{2\left(s\Delta_{SO}+\tau\Delta\right)}\right]\psi=\xi\psi.
\end{equation}
where $m_{eff}=\left|s\tau\Delta_{SO}+\Delta\right|/v_F^2$ is the electron
effective mass due to the band gap. This is the form of the Schr\"{o}dinger
equation in the presence of a magnetic field in which the emerging magnetic
moments are obviously manifested. Once again, we see the duality of the orbital
moments of the same nature as mentioned previously in the case of LLs: the
moments are coupled to SOC through spin and to mass through the valley degree
of freedom. Moreover, it is obvious that the expression for the magnetic moment is equal to the results of the low energy expansion given in Eq. \eqref{orbitalmagmom0}. Having in mind that these moments effectively shift the low energy parabolic bands, one can use the same argument as in Ref.~\onlinecite{cai13} to show that the separation between the lowest LL and the bottom of each {\it shifted} band is for each spin, valley and band to first order equal to half the separation between this and the first excited LL. This is in analogy with the LLs in a 2D massive-electron gas, where the lowest level sits at half the cyclotron frequency.\cite{cai13,koshino10} The difference for higher energy LLs is a consequence of the deviation of the dispersion from the quadratic one.

\section{Manifestation of orbital moments on magneto-transport}

We proceed with considering how the emerging magnetic moments affect the transport properties. In particular, we analyze transport through a single 1D barrier in
bulk graphene, extending from $x=0$ up to $x=W$, and along the $y$ direction, in which the intrinsic SOC is modified. The magnetic field is included only in the barrier, so we choose the following vector
potential (within the Landau gauge)
\begin{equation}\label{vecpot}
A_y=\begin{cases}0 &x<0\\Bx &0\leq x\leq W\\BW &x>W\end{cases}.
\end{equation}
The explicit derivation of the transmission coefficient is given in Appendix \ref{ap2}.

Since we analyze a barrier made exclusively out of SOC, the valley degree of
freedom plays no role in the electron transmission, which can be concluded from the theory presented in Secs. \ref{II} and \ref{III}. Therefore, the contour
plots of the transmission coefficient $T=\left|t\right|^2$, for the two spin
flavors, and for the $200$-nm wide barrier as a function of energy and the incident
angle of the incoming electron, are shown in Fig. 2. Each horizontal panel in
this figure corresponds to a specific value of the magnetic field, which is
$0$, $0.1$, $0.2$ to $0.3$ T from top to bottom. Because of the duality
$\Delta_{SO}\leftrightarrow\Delta$ and $s\leftrightarrow\tau$, the results
presented below also apply for transmission through a barrier when $\Delta\neq
0$ and $\Delta_{SO}=0$. But for this case the spin and valley quantum numbers
should be interchanged.

\begin{figure}
\centering
\includegraphics[width=8.6cm]{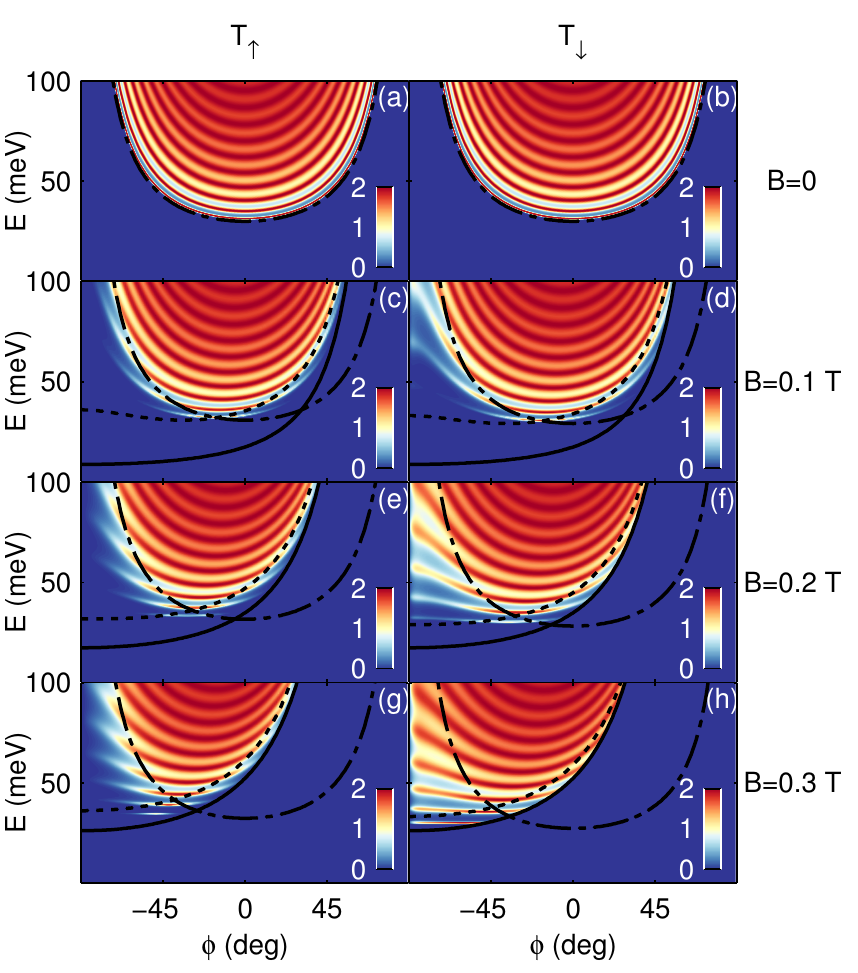}
\caption{Contour plots of the transmission coefficient as function of incident angle and energy for $\Delta_{SO}=30$ meV,
$\Delta=0$ and $W=200$ nm. The magnetic field equals $0$ T in (a-b), $0.1$ T in (c-d), $0.2$ T in (e-f), and $0.3$ T in (g-h).
The results are shown for both spin orientations. The semiclassical critical boundaries $\epsilon_{cr0}$ and
$\epsilon_{crW}$ are depicted by dash-dotted and dotted lines,
respectively.}
\label{1}
\end{figure}

For both barrier types, we found that the magnetic field causes cyclotron
motion, whose main feature is the appearance of a transmission window dependent on energy and incident angle $\phi$.\cite{masir08,grujic14} Outside of this window, the
waves after the barrier become evanescent, and therefore no transmission takes
place. This occurs when the longitudinal momentum
$k_x^{\prime}=\sqrt{\epsilon^2-k_y^{\prime}}$ of each electron state in the
region after the barrier becomes imaginary. The transmission window is given by
\begin{equation}\label{prozor}
\epsilon>\frac{\gamma}{1-\sin\phi},
\end{equation}
where $\gamma=W/l_B^2$. The transmission windows for different $B$ are shown by solid black curves in Fig. \ref{1}.

When the magnetic field increases, the transmission asymmetry with respect to
the incident angle becomes larger, due to the cyclotron motion, as shown in
Fig. \ref{1}. Besides, whereas transmission coefficients are identical for both spins when no magnetic field is present, $T_\uparrow$ and $T_\downarrow$ differ when $B\neq0$, which is a consequence of the SOC-induced magnetic moments. In
fact, it is clear from Eq. \eqref{difjna} that a quasi-classical longitudinal
momentum $q_x$
\begin{equation}\label{qx}
q_x\left(x\right)=\sqrt{\epsilon^2-\delta^2-\left(k_y+x/l_B^2\right)^2-s/l_B^2}
\end{equation}
can be assigned to the sublattice-polarized states.

In order to understand the effects of the emerging magnetic moments on the
transmission characteristics, it is instructive to investigate how classical
turning points vary with $\epsilon$ and $\phi$. Those turning points are
extracted from $q_x\left(x\right)=0$, where $q_x$ is given by Eq. \eqref{qx},
and are given by
\begin{equation}
x_{1,2}=-\epsilon l_B^2\sin\phi\mp l_B^2\sqrt{\epsilon^2-\delta^2-\mu},
\end{equation}
where $\mu=s/l_B^2$ is the magnetic moment term which appears in the expression for the quasi-classical momentum in Eq. \eqref{qx}. Given that the barrier
extends from $0$ to $W$, the condition that no turning points are found within
the barrier is obtained by requiring $x_1<0$ and $x_2>W$. The former condition
leads to
\begin{equation}
x_1<0\Rightarrow\begin{cases}\epsilon>\frac{\sqrt{\delta^2+\mu}}{\cos\phi},\quad\phi<0\\ \epsilon>\sqrt{\delta^2+\mu},\quad\phi>0\end{cases},
\end{equation}
while the latter results in
\begin{equation}
x_2>W\Rightarrow\begin{cases}\epsilon>\sqrt{\delta^2+\mu},\quad\epsilon\sin\phi+\gamma<0\\ \epsilon>\frac{\gamma\sin\phi+\sqrt{\gamma^2+\left(\delta^2+\mu\right)\cos^2\phi}}{\cos^2\phi},\quad\epsilon\sin\phi+\gamma>0\end{cases}.
\end{equation}

On the other hand, both classically forbidden and classically allowed regions
will be present in the barrier if $0<x_1<x_2<W$. The two extreme cases of
vanishing allowed regions occur when the leftmost turning point approaches the
right interface of the barrier
\begin{equation}
x_1<W\Rightarrow\begin{cases}\epsilon>\frac{\gamma\sin\phi+\sqrt{\gamma^2+\left(\delta^2+\mu\right)\cos^2\phi}}{\cos^2\phi},\quad\epsilon\sin\phi+\gamma<0\\ \epsilon>\sqrt{\delta^2+\mu},\quad\epsilon\sin\phi+\gamma>0\end{cases},
\end{equation}
and when the rightmost turning point approaches the left barrier interface
\begin{equation}
x_2>0\Rightarrow\begin{cases}\epsilon>\sqrt{\delta^2+\mu},\quad\phi<0\\ \epsilon>\frac{\sqrt{\delta^2+\mu}}{\cos\phi},\quad\phi>0\end{cases}.
\end{equation}
From the angle dependent functions in the last four equations one might define
the critical energies
\begin{equation}\label{ecr0}
\epsilon_{cr0}=\frac{\sqrt{\delta^2+\mu}}{\cos\phi},
\end{equation}
and
\begin{equation}\label{ecrW}
\epsilon_{crW}=\frac{\gamma\sin\phi+\sqrt{\gamma^2+\left(\delta^2+\mu\right)\cos^2\phi}}{\cos^2\phi},
\end{equation}
for which the classical turning points are located exactly at the two
interfaces, i.e. they are obtained by solving $q_x\left(0\right)=0$ and
$q_x\left(W\right)=0$, respectively. Those critical boundaries are plotted as
dash-dotted and dotted curves in Fig. \ref{1}.

\begin{figure}
\centering
\includegraphics[width=8.6cm]{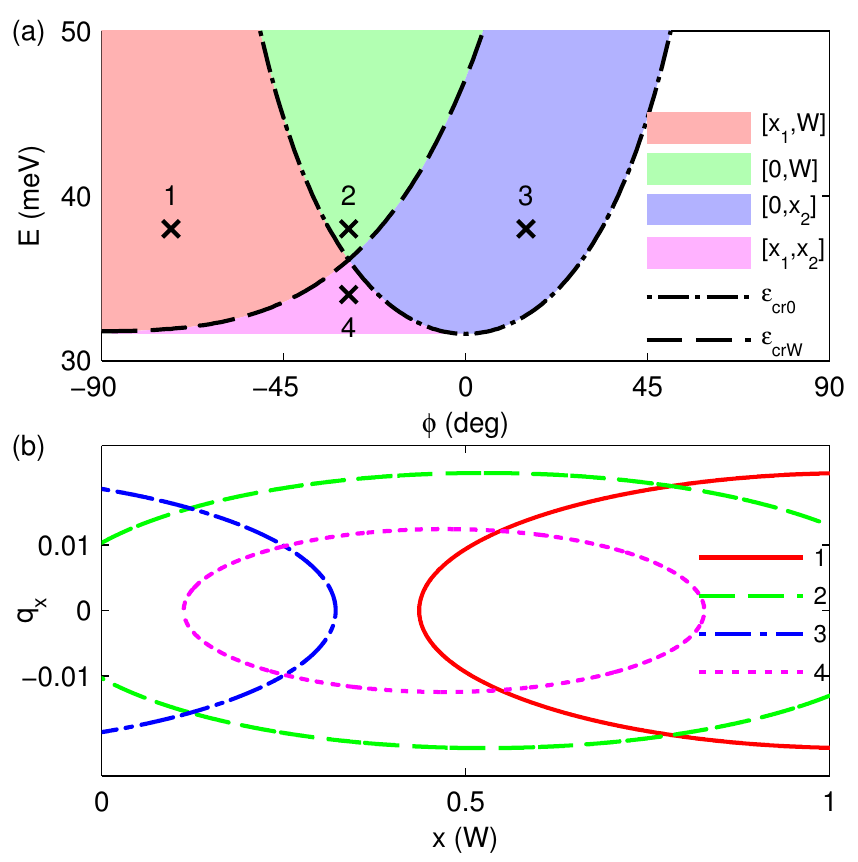}
\caption{(a) The regions with different ranges of turning points for $W=200$ nm, $\Delta_{SO}=30$ meV, $\Delta=0$ and $B=0.2$ T. Different classically allowed trajectories are found in differently shaded regions, demarcated by the two critical boundaries. (b) A family of four different classical trajectories which correspond to the states labeled by numbered crosses in each region  of (a).}
\label{zonekx}
\end{figure}

In order to elucidate the quasi-classical behavior, in Fig. \ref{zonekx}(a) we
plot the zones corresponding to different configurations of turning points by
different colors. The same set of parameters is used as in Fig. \ref{1}(e)
($\Delta_{SO}=30$ meV, $W=200$ nm, $B=0.2$ T and $s=+1$). In Fig.
\ref{zonekx}(b) we plot a set of classical trajectories that correspond to the zones shown in Fig. \ref{zonekx}(a). As could be inferred from Fig. 2, for
$\epsilon$ larger than both $\epsilon_{cr0}$ and $\epsilon_{crW}$ (green
colored region in Fig. \ref{zonekx}(a)), there is no classically forbidden
region inside the barrier. However, if the electron energy is between the two
critical energies (red or blue colored region in Fig. \ref{zonekx}(a)), a
classically forbidden energy range will appear on either end of the barrier. In other words, the electron will have to tunnel through a part of the barrier
adjacent to one of its interfaces, whereas propagation is free in the other
part. For the most extreme case displayed as the magenta colored region in Fig. \ref{zonekx}(a), the electron has to tunnel through both ends of the barrier.

One may notice that the two critical energies whose variation with $\phi$ is
depicted by dash-dotted and dotted lines in Fig. \ref{1} are almost identical
for the two spins. Also, by careful inspection of Fig. \ref{1} it becomes
evident that the quasi-classical zones we derived explain the observed
transmission very well, especially for the spin up states. For the spin down
states, however, transmission is enhanced with respect to the spin up states in the zones for which the electron waves must tunnel through a region of the
barrier (the red and blue energy zones in Fig. \ref{zonekx}(a)). This could be
understood if one recalls that the WKB expression for the tunneling coefficient
is given by
\begin{equation}
T\approx e^{-2Im\int q_x\left(x\right)dx},
\end{equation}
where the integration is over a classically forbidden region. Having this in
mind, it is obvious that for $\Delta_{SOC}\neq0$ and $B\neq0$ spin-up states
decay faster than the spin-down states in classically forbidden regions, due to the paramagnetic term. This difference increases at higher magnetic fields,
which leads to an increasing difference between the transmission coefficients
for the two spins, as Fig. \ref{1} clearly demonstrates. When the magnetic
field is absent, the emerging paramagnetism vanishes, and therefore, the
transmission characteristics for the two spins are identical (see Figs.
\ref{1}(a) and (b)).

Next, we explore how the presence of the magnetic moments affects the
interference pattern shown in Fig. \ref{1}. This could be the most important
effect from a practical point of view. In the Fabry-Perot model, the
interference pattern depends on the phase the electron wave function
accumulates between the barrier interfaces and the bounces from the
interface(s) and/or turning point(s)
\begin{equation}
\alpha=\alpha_{WKB}+\alpha_1+\alpha_2,
\end{equation}
where $\alpha_1$ and $\alpha_2$ are the backreflection phases, whereas
$\alpha_{WKB}$ is the WKB phase
\begin{equation}
\alpha_{WKB}=2\int_{\max\left(0,x_1\right)}^{\min\left(W,x_2\right)}q_x\left(x\right)dx.
\end{equation}
To analyze how the orbital magnetic moments influence the fringe pattern we
could once again invoke Eq. \eqref{qx} and the associated diagram in Fig.
\ref{zonekx}. It follows that Fabry-Perot resonances have different character
in the different zones. Whenever $B\neq0$, the WKB phase is accumulated
throughout the entire barrier for
$\epsilon>\max\left(\epsilon_{cr0},\epsilon_{crW}\right)$, but only in region
$\left[x_1,W\right]$ for $\epsilon_{cr0}>\epsilon>\epsilon_{crW}$ (the
red-shaded region in Fig. \ref{zonekx}(a)). Consequently, in the latter case
the transmission maxima (depicted by the red color in Fig. \ref{1}) are almost
linear functions of $\phi$, whereas in the former case their dependence on
$\phi$ is nonlinear.

The crucial point, however, is that the phase accumulated during the propagation
differs for the different spin orientations. This occurs because magnetic
moments associated with opposite spins contribute to $\alpha_{WKB}$ in opposite ways (see Eq. \eqref{qx}). To see this clearly, and to provide experimentally
verifiable predictions it is important to consider the conductivity of the
entire studied structure, given as,\cite{masir09}
\begin{equation}
G\left(\epsilon\right)=G_0\int_{-\pi/2}^{\pi/2}T\left(\epsilon,\phi\right)\epsilon\cos\phi d\phi,
\end{equation}
where $G_0=e^2L/2\hbar\pi^2$, with $L$ denoting the lateral width of the entire
structure in the $y$ direction.

\begin{figure}
\centering
\includegraphics[width=8.6cm]{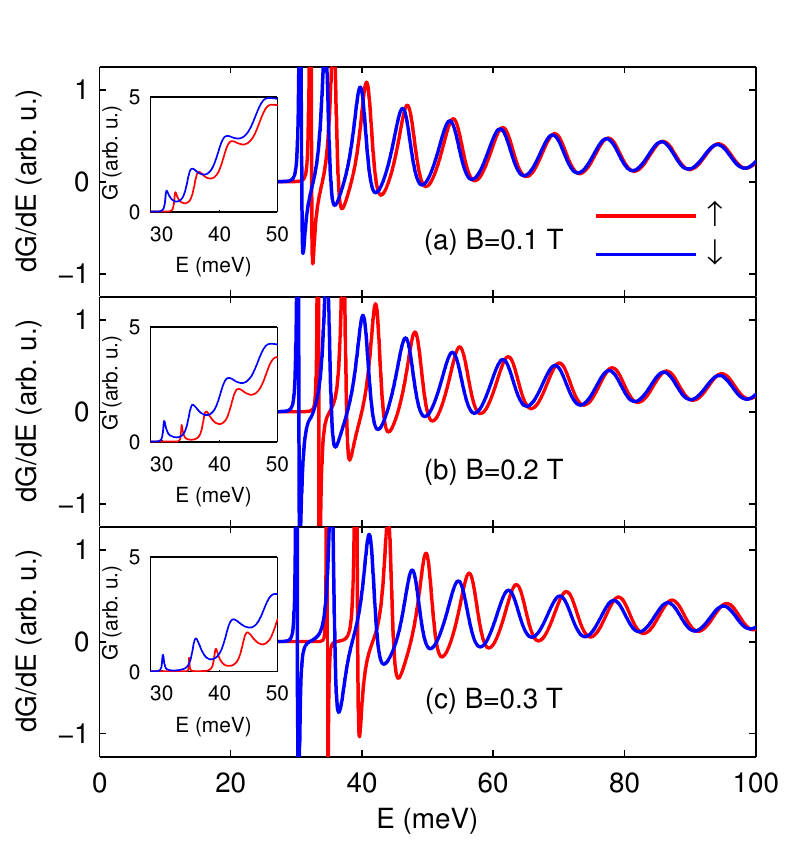}
\caption{The derivative of the conductance versus incident energy, for (a) $0.1$
T, (b) $0.2$ T and (c) $0.3$ T. All other parameters are the same as in Fig.
\ref{1}. Insets show the variation of the conductance with incident energy for
the corresponding magnetic field.}
\label{dGdE}
\end{figure}

Since the effects of magnetic moments are most vividly manifested in the
dependence of $dG/dE$ on energy, we display this quantity in Fig. \ref{dGdE},
for the same set of parameters as in Fig. \ref{1}. Alongside with $dG/dE$, the
corresponding conductance is shown in the insets for each case. As can be seen
from these insets, $G$ only depicts the fact that the spin-down conductance is
increased with respect to the spin-up conductance, due to the enhanced
transmission through the classically forbidden regions, as already discussed.
On the other hand, the first derivative of the conductance with respect to
energy conveys the information of the interference pattern, where the effects
of the orbital moments are more transparent. Two issues are of importance here: (i) The difference between the two spins is clearly more pronounced at higher
magnetic fields. This happens because in such a case the orbital moments have a larger impact on the electron dynamics, as pointed out before. (ii) The
distinction between the two spins is more prominent at lower energies. This is
a consequence of the larger emerging orbital magnetic moments of the electrons
whose energies are close to the band edges than of more energetic electrons, as
Eq. \eqref{orbitalmagmom} and Fig. \ref{0}(a) demonstrate.

\begin{figure}
\centering
\includegraphics[width=8.6cm]{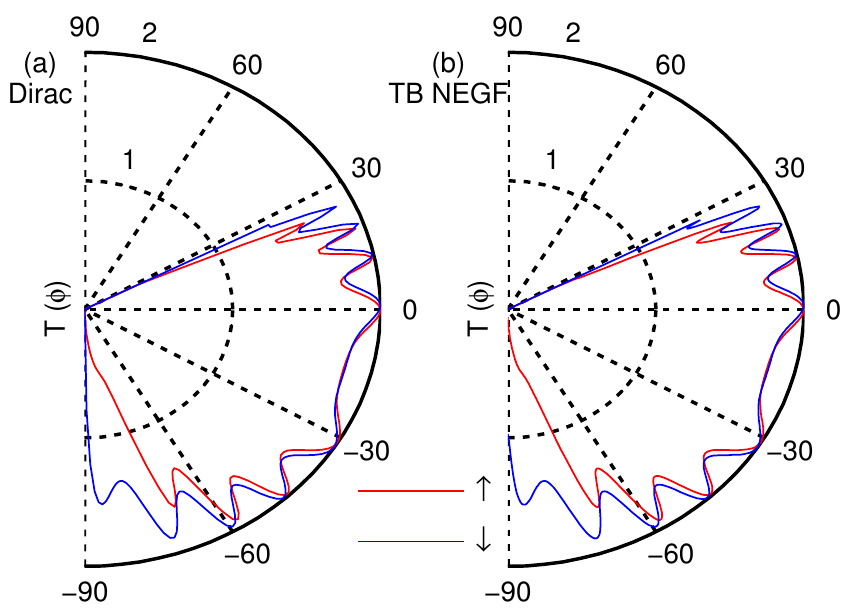}
\caption{Transmission curves calculated using (a) the continuum and (b) the TB
NEGF method. The parameters are $W=200$ nm, $B=0.3$ T and $E=100$ meV.}
\label{tbnegf}
\end{figure}

Finally, we would like to point out that the manifestation of orbital moments
in transport properties can be captured by the tight-binding nonequilibrium
Green function formalism as well. To show this, in Fig. \ref{tbnegf}(a) we plot a set of transmission curves obtained using the derived transmission amplitude, while in Fig. \ref{tbnegf}(b) we plot the results of
our numerical transport simulations within the TB NEGF method, for the same
barrier parameters. The phenomenological model used to describe graphene in this case is given by

\begin{equation}\label{eq:Hamiltonian}
H=-t\sum_{\langle i,j\rangle,\alpha}e^{i\varphi_{ij}}c^{\dagger}_{i\alpha}c_{j\alpha}+
i\lambda_{SO}\sum_{\langle\langle i,j\rangle\rangle,\alpha,\beta}\nu_{ij}e^{i\varphi_{ij}}c^{\dagger}_{i\alpha}s^z_{\alpha\beta}c_{j\beta}.
\end{equation}

The first term describes the usual hopping between nearest neighbor $p_z$
orbitals in graphene, which extends beyond the barrier. The second term
describes the intrinsic spin-orbit interaction found in the barrier, through
the next-nearest-neighbor (NNN) hopping amplitude $\lambda_{SO}$
($\Delta_{SO}=3\sqrt{3}\lambda_{SO}$). Note that $\nu_{ij}$ determines the sign of the hopping; it is positive (negative) if an electron makes a right (left) turn at the intermediate atom in hopping from site $j$ to site $i$. The Peierls term $\varphi_{ij}=\frac{e}{\hbar}\int_{\mathbf{r_i}}^{\mathbf{r_j}}\mathbf{A}\cdot
d\mathbf{l}$ accounts for the phase the electron acquires while traveling in
the presence of the magnetic field. The details of the NEGF procedure can be
found in Refs. ~\onlinecite{li08,datta95,futnota,liu12}.

Although the continuum and the TB NEGF schemes differ substantially as far as
the formalism and implementation are concerned, they give practically
indistinguishable results. This is not surprising, having in mind that the
continuum Dirac picture is the effective theory corresponding to the low energy tight-binding method. Therefore, both approaches display these Zeeman-type
effects, even though we use only minimal coupling and the Peierls substitution, to account for the influence of the magnetic field. Since strain in honeycomb lattices effectively induces time-reversal invariant pseudomagnetic fields,\cite{guinea10} stretching of insulating Dirac monolayers will inevitable galvanize orbital moments as well.\cite{grujic14}

Note that the TB NEGF method could prove handy for studying the effects of disorder and imperfections on the manifestation of the spin-contrasting orbital moments. However, unlike the orbital moments coupled to the spin, the valley-contrasting orbital moments can not be distinguished by the TB NEGF transport simulations, since the contributions from the two valleys are inherently summed together, and cannot be separated. In this case, only the continuum calculations, where the valley degree of freedom is explicit, can elucidate the underlying physics.

\section{Conclusion}

In this paper, we addressed the orbital magnetic moments emerging from the topology of insulating Dirac systems, as well as their manifestation on transport characteristics. In particular, we first closely examined the moments coupled to the spin degree of freedom, arising due to strong spin-orbit coupling, and thus leading to the renormalization of the g-factor. Their duality with the valley-contrasting orbital moments found in honeycomb lattices with broken spatial symmetry is reviewed, alongside with the duality of the Landau spectrum, particularly manifested in the behavior of the zeroth Landau level.

After establishing that magnetic properties couple with $\Delta_{SO}$ and the spin quantum number on the one hand, and $\Delta$ and the valley quantum number on the other hand in an analogous fashion, we go on to explore the influence of the orbital magnetic moments on the transport
properties. In particular, we focused on the transmission through a single 1D barrier made
of artificially enhanced spin-orbit coupling in graphene. We have shown that
certain Zeeman-like magneto-transport signatures are a clear manifestation of the induced
moments. The conductance $G$ through the device for the two
spins start deviating from each other with increasing magnetic field. The effects of the moments on the fringe pattern of the transmission
coefficients are most clearly observed in the energy dependence of the
derivative of the conductance with respect to the electron energy $dG/dE$. This quantity reflects the increasing shifts in the interference maxima of opposite
spins with increasing magnetic field; they are largest near the band edges, and they decrease for larger energies due to the decrease of the orbital magnetic
moments themselves.

Because of the analogy between the mass and the SOC terms and the orbital moments they induce, the results presented here are also valid for valley transmission through a barrier with only $\Delta\neq0$. This, however, can not be captured by numerical techniques such as the TB NEGF method, which is only able to account for the spin degree of freedom, and the associated orbital moments. Nevertheless, this behavior should be present in real devices, even in the absence of a clearly observable transport gap, since the Berry curvature hot spot can extend over a wide energy range.

\begin{acknowledgments}This work was supported by the Ministry of Education, Science and Technological Development (Serbia), and Fonds Wetenschappelijk Onderzoek (Belgium).
\end{acknowledgments}

\appendix

\section{Transmission through a barrier in bulk graphene}\label{ap2}

The studied structure and the chosen gauge for the vector potential (Eq. \eqref{vecpot}) ensure translational invariance along the $y$ direction, so $k_y$ is a good quantum number and the solutions have the form
$\Psi(x,y)=\exp(ik_yy)\left(\psi_A(x),\psi_B(x)\right)^T$. The following
coupled system of differential equations, for the amplitudes on the two sublattices can then be obtained:
\begin{equation}\label{veza}
\left(\tau k_x\mp ik_y\mp i\frac{e}{\hbar}A_y\right)\psi_{B/A}\pm\delta\psi_{A/B}=\epsilon\psi_{A/B}.
\end{equation}
Reducing the coupled system to a set of two independent second order
differential equations leads to
\begin{equation}
\left[\partial_x^2\mp\tau\frac{ e}{\hbar}(\partial_xA_y)-(k_y+\frac{e}{\hbar}A_y)^2+\epsilon^2-\delta^2\right]\psi_{A/B}=0.
\end{equation}

Having in mind the form of the vector potential, the differential equation in
the barrier becomes
\begin{equation}
\left[\partial_x^2\mp\frac{\tau}{l_B^2}-(k_y+\frac{x}{l_B^2})^2+\epsilon^2-\delta^2\right]\psi_{A/B}=0.
\end{equation}
By using the transformation $z=\sqrt{2}\left(k_yl_B+x/l_B\right)$ the following equation is
obtained
\begin{equation}
\left[\partial_z^2+1/2-1/2\mp\tau\frac{1}{2}+\left(\epsilon^2-\delta^2\right)\frac{l_B^2}{2}-\frac{z^2}{4}\right]\psi_{A/B}=0,
\end{equation}
which is of the form of the parabolic cylinder (Webers) differential equation
\begin{equation}
y^{\prime\prime}+\left(\nu+\frac{1}{2}-\frac{z^2}{4}\right)y=0,
\end{equation}
whose solutions are given in terms of parabolic cylinder functions
\begin{equation}
y=C_1D_{\nu}(z)+C_2D_{\nu}(-z).
\end{equation}

Finally the solution for the first sublattice is given by
\begin{equation}\label{psi21}
\begin{split}
\psi_A=&C_1D_{\nu_A}\left[\sqrt{2}\left(k_yl_B+x/l_B\right)\right]\\
+&C_2D_{\nu_A}\left[-\sqrt{2}\left(k_yl_B+x/l_B\right)\right],
\end{split}
\end{equation}
where $\nu_A=\left(\epsilon^2-\delta^2\right)l_B^2/2-\tau/2-1/2$. For the other sublattice after employing the recurrence relations
\begin{equation}
\frac{\partial D_{\nu}(z)}{\partial_z}=\frac{1}{2}zD_{\nu}(z)-D_{\nu+1}(z),
\end{equation}
and the relationship \eqref{veza}, one obtains the following expression
\begin{equation}\label{psi22}
\begin{split}
\psi_B=&C_1gD_{\nu_B}\left[\sqrt{2}\left(k_yl_B+x/l_B\right)\right]\\
-&C_2gD_{\nu_B}\left[-\sqrt{2}\left(k_yl_B+x/l_B\right)\right],
\end{split}
\end{equation}
where $\nu_B=\left(\epsilon^2-\delta^2\right)l_B^2/2+\tau/2-1/2$, and
\begin{equation}
g=i\left[\frac{\sqrt{2}}{\left(\epsilon+\tau\delta\right)l_B}\right]^{\tau}.
\end{equation}

If the relation
\begin{equation}
D_{\nu}\left(z\right)=2^{-\nu/2}e^{-z^2/4}H_{\nu}\left(\frac{z}{\sqrt{2}}\right)
\end{equation}
is employed, the spinor multiplied by $C_1$ in Eqs. \eqref{psi21} and
\eqref{psi22} reduces to the solution \eqref{spinorn}, once the incident energy is equal to a particular Landau level, as could be expected.

The incident wave function is given by

\begin{equation}\label{incident}
\psi_I=e^{ik_xx}\left(\begin{array}{c}1\\
\tau e^{i\tau\phi}\end{array}\right)+re^{-ik_xx}\left(\begin{array}{c}1\\ \tau e^{i\tau(\pi-\phi)}\end{array}\right),
\end{equation}
where $\phi=\arctan k_y/k_x$.

Finally, in the third region the vector potential is a non-zero constant, and
employing the standard plane wave ansatz, the solution is given by
\begin{equation}\label{psi3}
\psi_{III}=t\sqrt{\frac{k_x}{k_x^{\prime}}}e^{ik_x^{\prime}x}\left(\begin{array}{c}1\\ \tau e^{i\tau\theta}\end{array}\right),
\end{equation}
with the energy of the plane wave given by
$\epsilon=\alpha\sqrt{k_x^{\prime2}+k_y^{\prime2}}$,
$k_x^{\prime}=\epsilon\cos\theta$, the effective transverse momentum after the
barrier $k_y^{\prime}=\epsilon\sin\theta=k_y+W/l_B^2$ and $\theta$ being the
angle of energy propagation, with respect to the direction transverse to the
barrier. The additional factor under the square root follows from current
conservation.\cite{allain11} Again by replacing the expression for the momenta
before and after the barrier, one obtains the effective law of refraction for a barrier of thickness $W$ with nonzero $\Delta$, $\Delta_{SO}$ and $B$ as
\begin{equation}\label{refr1}
\epsilon\sin\theta=\epsilon\sin\phi+W/l_B^2.
\end{equation}

The expressions for the wavefunctions in different regions, \eqref{incident},
\eqref{psi21}, \eqref{psi22}, and \eqref{psi3} are then matched at the
interfaces $x=0$ and $x=W$, which gives a system of equations, whose solution
yields the transmission amplitude $t$
\begin{equation}\label{tampl}
t=\frac{2g\tau\cos(\tau\phi)\left(G_A^+G_B^-+G_A^-G_B^+\right)}{e^{ik_x^{\prime}W}f}\sqrt{\frac{k_x^{\prime}}{k_x}},
\end{equation}
where
\begin{equation}
\begin{split}
f=g^2\left(F_B^+G_B^--F_B^-G_B^+\right)+e^{i\tau\left(\theta-\phi\right)}\left(F_A^+G_A^--F_A^-G_A^+\right)\\
+g\tau e^{i\tau\theta}\left(F_B^-G_A^++F_B^+G_A^-\right)+g\tau e^{-i\tau\phi}\left(F_A^+G_B^-+F_A^-G_B^+\right).
\end{split}
\end{equation}
Here the coefficients $F^{\pm}$ and $G^{\pm}$ are given by
\begin{align}
F^{\pm}_{A/B}&=D_{\nu_{A/B}}\left[\pm\sqrt{2}k_yl_B\right],\\
G^{\pm}_{A/B}&=D_{\nu_{A/B}}\left[\pm\sqrt{2}(k_yl_B+\frac{W}{l_B})\right].
\end{align}


\begin{thebibliography}{00}
\bibitem{footnote}$\Delta_{SO}$ also gives rise to massive particles, by virtue of opening the gap, however for brevity we refer only to the $\Delta$ term as mass.
\bibitem{futnota}In order to compare the results with the continuum theory, we
    need
    the
    numerical simulations within the NEGF formalism to provide us with the
    angular dependence of the transmission through a structure infinite in the
    $y$-direction. To achieve this, we resort to the recipe described in
    Ref.~\onlinecite{liu12}. In short, we take the narrowest possible zigzag
    nanoribbon, placed along the $x$-direction, where the semi-infinite left
    and right leads surround the barrier region in which SOC NNN hopping is
    nonzero. The structure is then taken to be periodic along the
    $y$-direction, prompting the use of the Bloch theorem in this direction.
    This means that the phase factor will enter all hopping terms along the $y$
    axis, which is none other than the transverse momentum $k_y$. In this way,
    $k_y$ appears as a parameter, and since the incident energy $E$ is a
    parameter as well, one is then able to reconstruct the angle of propagation
    using $\hbar v_Fk_y=E\sin\phi$. Note that in our case, besides the Peierls
    phase factor, we must also add the vector potential, Eq. \eqref{vecpot}, to
    $k_y$, in order to properly model the influence of the magnetic field.
    Finally, one needs to connect the Fermi velocity $v_F$ entering the Dirac
    equation, with the nearest-neighbor hopping $t$, as $v_F=3ta/2\hbar$, where
    $a=0.142$ nm is the carbon-carbon distance, and the hopping is set to
    $t=2.7$ eV.
\end{thebibliography}
\end{document}